\documentclass[preprint,preprintnumbers, prd, floatfix, superscriptaddress,nofootinbib] {revtex4-1}
\usepackage{epsfig}
\usepackage{subfigure}
\usepackage{dcolumn}
\usepackage{bm}
\usepackage[usenames ,dvipsnames]{xcolor}
\usepackage{slashed}
\usepackage{graphicx,color}

\begin{document}

\title{Baryonic $B$ meson decays}

\author{Xiaotao Huang}
\affiliation{The Institute for Advanced Studies, Wuhan University, Wuhan 430072, China}

\author{Yu-Kuo Hsiao}
\email{Corresponding author: yukuohsiao@gmail.com}
\affiliation{School of Physics and Information Engineering, Shanxi Normal University, Linfen 041004, China}

\author{Jike Wang}
\email{Corresponding author: Jike.Wang@whu.edu.cn}
\affiliation{The Institute for Advanced Studies, Wuhan University, Wuhan 430072, China}

\author{Liang Sun}
\email{Corresponding author: sunl@whu.edu.cn}
\affiliation{School of Physics and Technology, Wuhan University, Wuhan 430072, China}

\date{\today}

\begin{abstract}
We review the two and three-body baryonic $B$ decays 
with the dibaryon (${\bf B\bar B'}$) as the final states. Accordingly,
we summarize the experimental data
of the branching fractions, angular asymmetries, and $CP$ asymmetries.
Using the $W$-boson annihilation (exchange) mechanism,
the branching fractions of $B\to {\bf B \bf \bar B'}$ are shown to be interpretable.
In the approach of perturbative QCD counting rules,
we study the three-body decay channels.
In particular, we review the $CP$ asymmetries of $B\to {\bf B\bar B'}M$,
which are promising to be measured by the LHCb and Belle~II experiments.
Finally, we remark the theoretical challenges in interpreting 
${\cal B}(B^-\to p\bar p\rho^-)$ and ${\cal B}(B^-\to p\bar p\mu^-\bar \nu_\mu)$.
\end{abstract}

\maketitle
\section{Introduction}
The baryonic $B$ meson decays have been richly measured with 
the branching fractions, angular asymmetries, and $CP$ asymmetries
in two and three-body decay channels~\cite{Abe:2002ds,LHCb:2014nix,Aubert:2005gw,
Aubert:2006qx,Chen:2008jy,Wei:2007fg,Aaij:2013fla,Aaij:2017vnw,
Aaij:2017gum,Aaij:2016xfa,Aaij:2013fla,pdg,Belle:2007lbz,Chang:2015fja}, 
as summarized in Table~\ref{data1}.
Typically, ${\cal B}(B\to{\bf B\bar B'})$ is as small as $10^{-8}-10^{-7}$.
Nonetheless, it is observed that
${\cal B}(B\to{\bf B\bar B'}M)\sim (10-100)\times {\cal B}(B\to{\bf B\bar B'})$,
due to a sharply rising peak in $B\to{\bf B\bar B'}M$
observed around the threshold area of $m_{\bf B\bar B'}\sim m_{\bf B}+m_{\bf \bar B'}$
in the dibaryon invariant mass spectrum~\cite{Aubert:2005gw}.
Known as the threshold effect, 
it enhances ${\cal B}(B\to{\bf B\bar B'}M)$ as large as $10^{-6}$.
While the $\bf B\bar B'$ production shows the tendancy to occur around 
$m_{\bf B\bar B'}\sim m_{\bf B}+m_{\bf B'}$,
$B\to{\bf B\bar B'}$ proceeds at $m_B$ scale, far from the threshold area.
This interprets the suppressed ${\cal B}(B\to{\bf B\bar B'})$~\cite{Hou:2000bz,Suzuki:2006nn}.

The partial branching fraction of $B\to{\bf B\bar B'}M$ can be a function of $\cos\theta_{\bf B}$, 
where $\theta_{\bf B}$ is the angle between the baryon and meson moving directions
in the dibaryon rest frame.
One hence defines the forward-backward angular asymmetry,
\begin{eqnarray}
{\cal A}_{FB}\equiv \frac
{{\cal B}(\cos\theta_{\bf B}>0)-{\cal B}(\cos\theta_{\bf B}<0)}
{{\cal B}(\cos\theta_{\bf B}>0)+{\cal B}(\cos\theta_{\bf B}<0)}\,.
\end{eqnarray}
In Table~\ref{data1},
${\cal A}_{FB}(B^-\to p\bar p \pi^{-},p\bar p K^-)$
$=(-40.9\pm 3.4,49.5\pm 1.4)\%$~\cite{LHCb:2014nix}
indicate that one of the dibaryon favors to move collinearly with the meson.

We search for the theoretical approach to interpret
the threshold effect, branching fractions and angular asymmetries
of the baryonic $B$ decays. We find that
the factorization approach can be useful~\cite{ali}, where
one factorizes (decomposes) the amplitude of the decay as two separate matrix elements. 
In our case, we present 
\begin{eqnarray}
{\cal M}(B\to{\bf B\bar B'})&\propto& 
\langle {\bf B\bar B'}|(\bar q q')|0\rangle \langle 0|(\bar q b)|B\rangle\,,\nonumber\\
{\cal M}_1(B\to{\bf B\bar B'}M)&\propto&
\langle {\bf B\bar B'}|(\bar q q')|0\rangle \langle M|(\bar q b)|B\rangle\,,\nonumber\\
{\cal M}_2(B\to{\bf B\bar B'}M)&\propto&
\langle M|(\bar q q')|0\rangle \langle {\bf B\bar B'}|(\bar q b)|B\rangle\,,
\end{eqnarray}
where $(\bar q q')$ and $(\bar q b)$ stand for the quark currents,
and the matrix element of $\langle {\bf B\bar B'}|(\bar q q')|0\rangle$
($\langle {\bf B\bar B'}|(\bar q b|B\rangle$) can be parameterized as 
the timelike baryonic ($B$ to ${\bf B\bar B'}$ transition) form factors $F_{\bf B\bar B'}$.
Moreover, one derives $F_{\bf B\bar B'}\propto 1/t^n$ 
in perturbative QCD (pQCD) counting rules~\cite{Brodsky:1973kr,Brodsky:1974vy,Lepage:1979za,
Lepage:1980fj,Brodsky:1973kr,Brodsky:2003gs,Chua:2002wn,Geng:2006wz,Chua:2001vh},
where $t\equiv (p_{\bf B}+p_{\bf  \bar B'})^2$ and
$n$ accounts for the number of the gluon propagators that attach to the baryon pair.
It results in $d{\cal B}/dm_{\bf B\bar B'}\propto1/t^{2n}$, 
which shapes a peak around $m_{\bf B\bar B'}\sim m_{\bf B}+m_{\bf B'}$ 
in the $m_{\bf B\bar B'}$ spectrum, and then the threshold effect can be interpreted.
In the $B\to p\bar p$ transition,
there exists the term of $(p_{\bar p}-p_{p})_\mu\bar u(\gamma_5)v$ 
for $F_{\bf B\bar B'}$~\cite{Geng:2006wz},
which is reduced as $(E_{\bar p}-E_{p})\bar u(\gamma_5)v$ in the $p\bar p$ rest frame. 
Since $(E_{\bar p}-E_{p})\propto\cos\theta_p$, 
the term for $F_{\bf B\bar B'}$ can be used to describe the highly asymmetric 
${\cal A}_{FB}(B^-\to p\bar p \pi^{-},p\bar p K^-)$.
Alternatively, the baryonic $B$ decays is studied 
with the pole model, where the non-factorizable contributions
can be taken into account~\cite{HYppD,HYpole,HYchamBaryon,HYreview}.

%
\begin{table}[b!]
\caption{The measured branching fractions, forward-backward asymmetries (${\cal A}_{FB}$), and 
$CP$ asymmetries (${\cal A}_{CP}$) for the baryonic $B$ decays, where
the notation $\dagger$ is for ${\cal A}_{FB}$ with $m_{p\bar p}<2.85$~GeV.}\label{data1}
{
\tiny
\begin{tabular}{|l|rcc|l|}
\hline
Decay mode              &Branching fraction&${\cal A}_{FB}$&${\cal A}_{CP}$&Ref.\\
\hline\hline
$\bar B^0\to p\bar p$ 
&$(1.25\pm 0.32)\times 10^{-8}$
&&
&\cite{pdg}\\

$\bar B^0\to \Lambda\bar \Lambda$ 
&$<3.2\times 10^{-7}$
&&
&\cite{pdg}\\

$B^-\to \Lambda\bar p$ 
&$(2.4^{+1.0}_{-0.9})\times 10^{-7}$
&
&
&\cite{pdg}\\

$\bar B_s^0\to p\bar p$
& $<1.5\times 10^{-8}$
&&
&\cite{pdg}\\
\hline

$\bar B^0\to p\bar p\pi^0$
&$(5.0\pm 1.9)\times 10^{-7}$
&&
&\cite{pdg}\\

$\bar B^0\to p\bar p \bar K^{0}$
&$(2.66\pm 0.32)\times 10^{-6}$
&
&
&\cite{pdg}\\

$\bar B^0\to \Lambda\bar p\pi^+$
&$(3.14\pm 0.29)\times 10^{-6}$
&$-0.41\pm 0.11\pm 0.03$
&$0.04\pm 0.07$
&\cite{pdg,Belle:2007lbz}\\

$\bar B^0\to \Sigma^0\bar p\pi^+$
&$<3.8\times 10^{-6}$
&&
&\cite{pdg}\\

$\bar B^0\to \Lambda\bar p K^+$
&$<8.2\times 10^{-7}$
&&
&\cite{pdg}\\

$\bar B^0\to \Lambda\bar \Lambda \bar K^{0}$
&$(4.8^{+1.0}_{-0.9})\times 10^{-6}$
&&
&\cite{pdg}\\

$B^-\to p\bar p \pi^{-}$
&$(1.62\pm 0.20)\times 10^{-6}$
&$(-0.409\pm 0.033\pm 0.006)^\dagger$
&$0.00\pm 0.04$
&\cite{pdg,LHCb:2014nix}\\

$B^-\to p\bar p K^{-}$
&$(5.9\pm 0.5)\times 10^{-6}$
&$(0.495\pm 0.012\pm 0.007)^\dagger$
&$0.00\pm 0.04$
&\cite{pdg,LHCb:2014nix}\\

$B^-\to \Lambda\bar p \pi^0$
&$(3.0^{+0.7}_{-0.6})\times 10^{-6}$
&$-0.16\pm 0.18\pm 0.03$
&$0.01\pm 0.17$
&\cite{pdg,Belle:2007lbz}\\

$B^-\to \Lambda\bar \Lambda \pi^-$
&$<9.4\times 10^{-7}$
&&
&\cite{pdg}\\

$B^-\to \Lambda\bar \Lambda K^{-}$
&$(3.4\pm 0.6)\times 10^{-6}$
&&
&\cite{pdg}\\

$\bar B^0_s\to \bar p \Lambda K^+$+c.c.
&$(5.5\pm 1.0)\times 10^{-6}$
&&
&\cite{pdg}\\

\hline
$\bar B^0\to p\bar p \bar K^{*0}$
&$(1.24^{+0.28}_{-0.25})\times 10^{-6}$
&
&$0.05\pm 0.12$
&\cite{pdg}\\

$\bar B^0\to \Lambda\bar \Lambda \bar K^{*0}$
&$(2.5^{+0.9}_{-0.8})\times 10^{-6}$
&&
&\cite{pdg}\\

$B^-\to p\bar p K^{*-}$
&$(3.6^{+0.8}_{-0.7})\times 10^{-6}$
&
&$0.21\pm 0.16$
&\cite{pdg}\\

$B^-\to \Lambda\bar p \rho^0,\rho^0\to\pi^+\pi^-$
&$(4.8\pm 0.9)\times 10^{-6}$
&&
&\cite{pdg}\\

$B^-\to \Lambda\bar p \phi$
&$(8.0\pm 2.2)\times 10^{-7}$
&&
&\cite{pdg}\\

$B^-\to \Lambda\bar \Lambda K^{*-}$
&$(2.2^{+1.2}_{-0.9})\times 10^{-6}$
&&
&\cite{pdg}\\
\hline
\hline
$\bar B^0\to \Lambda_c^+ \bar p$
&$(1.54\pm 0.18)\times 10^{-5}$
&&
&\cite{pdg}\\
$\bar B^0\to \Sigma_c^+ \bar p$
&$<2.4\times 10^{-5}$
&&
&\cite{pdg}\\
$B^-\to \Sigma_c^0 \bar p$
&$(2.9\pm 0.7)\times 10^{-5}$
&&
&\cite{pdg}\\
\hline
$B^-\to p\bar p D^-$
&$<1.5\times 10^{-5}$
&&
&\cite{pdg}\\
$B^-\to p\bar p D^{*-}$
&$<1.5\times 10^{-5}$
&&
&\cite{pdg}\\
$B^-\to \Lambda\bar p D^0$
&$(1.43\pm 0.32)\times 10^{-5}$
&&
&\cite{pdg}\\
$B^-\to \Lambda\bar p D^{*0}$
&$<5\times 10^{-5}$
&&
&\cite{pdg}\\
$\bar B^0\to n\bar p D^{*+}$
&$(1.4\pm 0.4)\times 10^{-3}$
&&
&\cite{pdg}\\

$\bar B^0\to p\bar p D^{0}$
&$(1.04\pm 0.07)\times 10^{-4}$
&&
&\cite{pdg}\\
$\bar B^0\to p\bar p D^{*0}$
&$(0.99\pm 0.11)\times 10^{-4}$
&&
&\cite{pdg}\\

$\bar B^0\to \Lambda\bar p D_s^+$
&$(2.8\pm 0.9)\times 10^{-5}$
&
&
&\cite{pdg}\\

$\bar B^0\to \Lambda\bar p D^+$
&$(2.5\pm 0.4)\times 10^{-5}$
&$-0.08\pm 0.10$
&
&\cite{pdg,Chang:2015fja}\\

$\bar B^0\to \Lambda\bar p D^{*+}$
&$(3.4\pm 0.8)\times 10^{-5}$
&$+0.55\pm 0.17$
&
&\cite{pdg,Chang:2015fja}\\

$\bar B^0\to \Lambda\bar \Lambda \bar D^0$
&$(1.00^{+0.30}_{-0.26})\times 10^{-5}$
&
&
&\cite{pdg}\\

$\bar B^0\to \Sigma^0\bar \Lambda \bar D^0$+c.c.
&$<3.1\times 10^{-5}$
&
&
&\cite{pdg}\\
\hline
\end{tabular}}
\end{table}
%
We have explained ${\cal B}(B\to{\bf B\bar B'})\sim 10^{-8}-10^{-7}$~\cite{Hsiao:2014zza}.
We have studied $B\to{\bf B\bar B'}M$,
and explained the branching fractions and $CP$ asymmetries~\cite{Geng:2006jt,Hsiao:2019ann,
Geng:2005fh,Hsiao:2015mca,Chen:2008sw,Hsiao:2016amt,
Geng:2011pw,Geng:2016fdw,Geng:2007cw}. 
In addition, we have predicted
${\cal B}(\bar B^0_s\to p\bar \Lambda K^- + \Lambda\bar p K^+)
=(5.1\pm 1.1)\times 10^{-6}$~\cite{Geng:2016fdw},
in excellent agreement with the value of
$(5.46\pm 0.61\pm 0.57\pm 0.50\pm 0.32)\times 10^{-6}$ measured
by LHCb~\cite{Aaij:2017vnw}. This demonstrates that 
the theoretical approach can be reliable.
Therefore, we would like to present a review, 
in order to illustrate how the approach of pQCD counting rules
based on the factorization can be applied to the baryonic $B$ decays. 
We will also review our theoretical results that have explained the branching fractions of 
$B\to{\bf B\bar B'}$ and $B\to{\bf B\bar B'}M$; particularly, the $CP$ asymmetries, 
promising to be observed by future measurements.

\section{Formalism}
%
\begin{figure}[t!]
\includegraphics[width=3.8in]{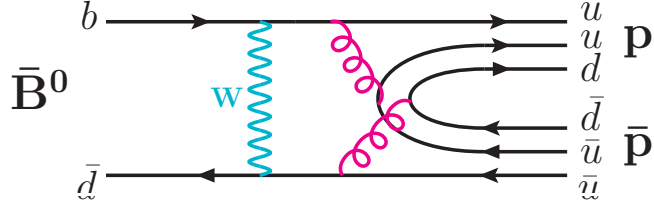}
\caption{Feynman diagram for $\bar B^0\to p\bar p$.}\label{fig1}
\end{figure}
%
To review the two-body baryonic $B$ decays, 
we take $\bar B^0\to p\bar p$ as our example. 
According to Fig.~\ref{fig1}, $\bar B^0\to p\bar p$ is regarded 
as an $W$-boson exchange process~\cite{Bediaga:1991eu,Pham:1980dc,
Pham:1980xe,Hsiao:2014zza}. In the factorization,
we derive the amplitude as~\cite{Hsiao:2014zza}
\begin{eqnarray}\label{amp_2b}
&&{\cal M}(\bar B^0\to p\bar p)=\frac{G_F}{\sqrt 2}
V_{ub}V_{ud}^*\,a_2\langle p\bar p|\bar u\gamma_\mu(1-\gamma_5) u|0\rangle
\langle 0|\bar d\gamma^\mu(1-\gamma_5) b|\bar B^0\rangle\,,
\end{eqnarray}
where $G_F$ is the Fermi constant, and $V_{ub(d)}^{(*)}$ 
the Cabibbo-Kobayashi-Maskawa (CKM) matrix element.
One has defined $\langle 0|\bar d\gamma^\mu(1-\gamma_5) b|\bar B^0\rangle=-if_Bq_\mu$
for the $\bar B^0$ meson annihilation, where $f_B$ is the decay constant and $q_\mu$
the four-momentum.
For the $p\bar p$ production, the matrix elements read~\cite{Chua:2002wn,Chua:2001vh}
\begin{eqnarray}\label{timelikeF}
\langle {\bf B\bar B'}|V_\mu|0\rangle
&=&\bar u\bigg[F_1\gamma_\mu+\frac{F_2}{m_{\bf B}+m_{\bf\bar B'}}i\sigma_{\mu\nu}q_\mu\bigg]v\;,\nonumber\\
\langle {\bf B\bar B'}|A_\mu|0\rangle
&=&\bar u\bigg[g_A\gamma_\mu+\frac{h_A}{m_{p}+m_{\bar p}}q_\mu\bigg]\gamma_5 v\,,
\end{eqnarray}
with the (axial-)vector current $V(A)_\mu=\bar q\gamma_\mu(\gamma_5) q'$,
where $F_{1,2}$, $g_A$ and $h_A$ are the timelike baryonic form factors.

\newpage
At a very large momentum transfer ($Q^2\to\infty$),
the approach of pQCD counting rules
results in~\cite{Brodsky:1973kr,Brodsky:1974vy,Lepage:1979za,Lepage:1980fj,Brodsky:2003gs}
\begin{eqnarray}\label{F1gA1}
F_1,g_A\propto \frac{\alpha_s^2(Q^2)}{Q^4}
\ln\bigg(\frac{Q^2}{\Lambda^2}\bigg)^{-\frac{4}{3\beta}}\,,
\end{eqnarray}
where $\beta=11-2n_f/3$ is the $\beta$ function of QCD to one loop,
$n_f=3$ the flavor number, and $\Lambda=0.3$~GeV the scale factor. 
Moreover, $\alpha_s(Q^2)\equiv (4\pi/\beta)[\text{ln}(Q^2/\Lambda^2)]^{-1}$
is the running coupling constant in the strong interaction~\cite{Lepage:1980fj}.
Interestingly, $\alpha_s^2/Q^4$ reflects the fact that 
one needs two hard gluon propagators to attach to the baryons
as drawn in Fig.~\ref{fig1}~\cite{E835:1999mlt}, whereas 
$\ln(Q^2/\Lambda^2)^{-4/(3\beta)}$ is caused by the wave function.

As $V_\mu$ and $A_\mu$ are combined as 
the right or left-handed chiral current, that is, $J^{R,L}_\mu=(V_\mu\pm A_\mu)/2$, 
one obtains $\langle {\bf B}'_{R+L}|J^{R(L)}_\mu|{\bf B}_{R+L}\rangle$ 
for the spacelike ${\bf B}\to{\bf B'}$ transition.
With the right-handed current, the matrix elements 
can be written as~\cite{Lepage:1979za,Hsiao:2014zza}
\begin{eqnarray}\label{Gff1}
\langle {\bf B'}_{R+L}|J_\mu^R|{\bf B}_{R+L}\rangle=
\bar u\bigg[\gamma_\mu \frac{1+\gamma_5}{2}F_R+\gamma_\mu \frac{1-\gamma_5}{2}F_L\bigg]u\,,
\end{eqnarray}
where $|{\bf B}_{R+L}\rangle=|{\bf B}_R\rangle+|{\bf B}_L\rangle$,
and $F_{R,L}$ are the chiral form factors. 
With $q_i$ ($i=1,2,3$) denoting one of the valence quarks in ${\bf B}$,
$Q\equiv J^R_{\mu=0}$ known as the chiral charge  
is able to change the flavor for $q_i$, such that ${\bf B}$ is transformed as ${\bf B'}$.
Note that the chirality is regarded as the helicity at $Q^2\to\infty$.
Since the helicity of $q_i$ can be (anti-)parallel $[||(\overline{||})]$ to the helicity of ${\bf B}$,
we define $Q_{||(\overline{||})}(i)$ that is responsible for acting on $q_i$.
Thus, the approach of pQCD counting rules lead to~\cite{Lepage:1979za,Hsiao:2014zza}
\begin{eqnarray}\label{Gff2}
F_R=e^R_{||}F_{||}+e^R_{\overline{||}}F_{\overline{||}}\,,\;\;
F_L=e^L_{||}F_{||}+e^L_{\overline{||}}F_{\overline{||}}\,,
\end{eqnarray}
with $e^R_{||(\overline{||})}=\langle {\bf B'}_R|{\bf Q_{||(\overline{||})}}|{\bf B}_R\rangle$,
$e^L_{||(\overline{||})}           =\langle {\bf B'}_L|{\bf Q_{||(\overline{||})}}|{\bf B}_L\rangle$
and ${\bf Q_{||(\overline{||})}}=\sum_i Q_{||(\overline{||})}(i)$, where
the $SU(3)$ flavor ($SU(3)_f$) and $SU(2)$ spin symmetries are both respected.
In the crossing symmetry, the spacelike form factors behave as the timelike ones, 
such that one can relate $F_1$ and $g_A$ with the chiral form factors
in Eqs.~(\ref{Gff1},\,\ref{Gff2}) derived in the spacelike region,
leading to $F_1(g_A)=(e^R_{||}\pm e^L_{||})F_{||}
+(e^R_{\overline{||}}\pm e^L_{\overline{||}})F_{\overline{||}}$.
In addition to the momentum dependence of Eq.~(\ref{F1gA1}),
$F_1$ and $g_A$ are 
presented as~\cite{Hsiao:2014zza,Chua:2001vh,Chua:2002wn}
\begin{eqnarray}
&&
F_1=\frac{C_{F_1}}{t^2}\ln\bigg(\frac{t}{\Lambda^2}\bigg)^{-\gamma}\,,\;
g_A=\frac{C_{g_A}}{t^2}\ln\bigg(\frac{t}{\Lambda^2}\bigg)^{-\gamma}\,,
\end{eqnarray}
where $\gamma= 2+4/(3\beta)=2.148$.

\newpage
For $\langle p\bar p|(\bar u u)|0\rangle$,
we obtain~\cite{Chua:2002wn}
\begin{eqnarray}\label{0toBB}
C_{F_1}=\frac{5}{3}C_{||}+\frac{1}{3}C_{\overline{||}}\,,
C_{g_A}=\frac{5}{3}C_{||}-\frac{1}{3}C_{\overline{||}}\,,
\end{eqnarray}
where $C_{||(\overline{||})}$ is from 
$F_{||(\overline{||})}\equiv C_{||(\overline{||})}/t^2[\ln(t/\Lambda^2)]^{-\gamma}$,
and we have used $(e^R_{||},e^L_{||})=(5/3,0)$ and 
$(e^R_{\overline{||}},e^L_{\overline{||}})=(0,1/3)$~\cite{Chua:2002wn,Hsiao:2014zza}.
In Ref.~\cite{Belitsky:2002kj}, the pQCQ calculation causes $F_2=F_1/(t\text{ln}[t/\Lambda^2])$,
indicating that $F_2$ has a suppressed contribution. Taking the form factors as the inputs,
we reduce the amplitude of $\bar B^0\to p\bar p$ as
\begin{eqnarray}\label{A1A2}
&&{\cal M}\propto
\frac{1}{(m_{p}+m_{\bar p})}\;\bar u\bigg[(m_{p}+m_{\bar p})^2 g_A
+m_{B}^2 h_A\bigg]\gamma_5 v\,,
\end{eqnarray}
where $F_{1(2)}$ has been vanishing,
in accordance with the conservation of vector current (CVC)
$q^\mu\langle p\bar p|\bar u\gamma_\mu u|0\rangle=0$. 
Using the partial conservation of axial-vector current (PCAC),
where $q^\mu \langle {\bf B\bar B'}|A_\mu|0\rangle=0$,
it is obtained that~\cite{Chua:2002wn,Chen:2008pf,
Bediaga:1991eu,Pham:1980dc,Pham:1980xe,Hsiao:2014zza},
\begin{eqnarray}\label{hA}
h_A=-\frac{(m_{\bf B}+m_{\bf B'})^2 g_A}{t-m_{M}^2}\,,
\end{eqnarray}
by which $g_A$ and $h_A$ cancel each other, and then
${\cal M}\simeq 0$. This seems that  
the $W$-exchange (annihilation) mechanism 
based on the factorization fails to explain ${\cal B}(B\to{\bf B\bar B'})$.
As a consequence, one turns to think of the non-factorizable effects 
as the main contributions~\cite{Chernyak:1990ag,Ball:1990fw,
Chang:2001jt,
HYpole,Chua:2013zga,He:2006vz,Cheng:2002sa,Chen:2008pf}.

In Eq.~(\ref{hA}), $1/(t-m_{M}^2)$ describes a meson pole, so that
$B\to{\bf B\bar B'}$ can be regarded to receive the contribution
from the intermediate process of $B\to M\to {\bf B\bar B'}$,
which is much suppressed. On the other hand, 
there might exist a QCD-based contribution to $h_A$, by which
${\cal M}(B\to {\bf B\bar B'})\neq 0$, and PCAC is violated. 
Here, we choose to parameterize $h_A$ with slightly violated PCAC in the timelike region.
To this end, we derive $h_A+g_A\simeq 0$ with $q^\mu \langle {\bf B\bar B'}|A_\mu|0\rangle\simeq 0$ 
at the threshold area of $t \simeq (m_{\bf B}+m_{\bf \bar B'})^2$,  
where the meson pole is supposed to be inapplicable~\cite{Pham:1980dc,Pham:1980xe}.
Since the QCD-based calculation of $h_A$ is still lacking,
besides $h_A+g_A\simeq 0$ suggests $h_A\propto g_A$, 
we are allowed to present $h_A=C_{h_A}/t^2$ for its momentum dependence~\cite{Hsiao:2014zza}.
%
\begin{figure}
\includegraphics[width=2.4in]{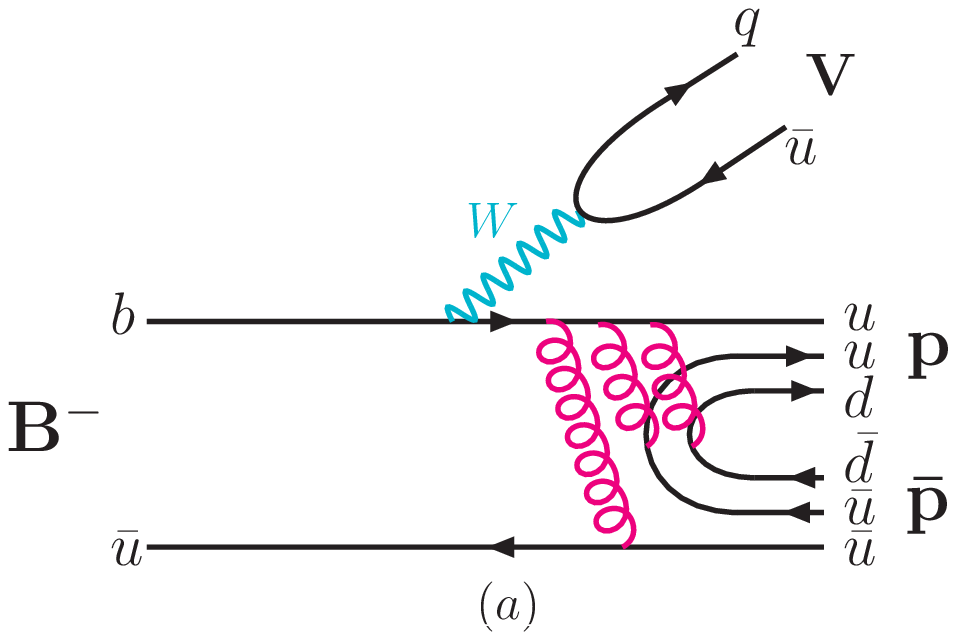}
\includegraphics[width=2.0in]{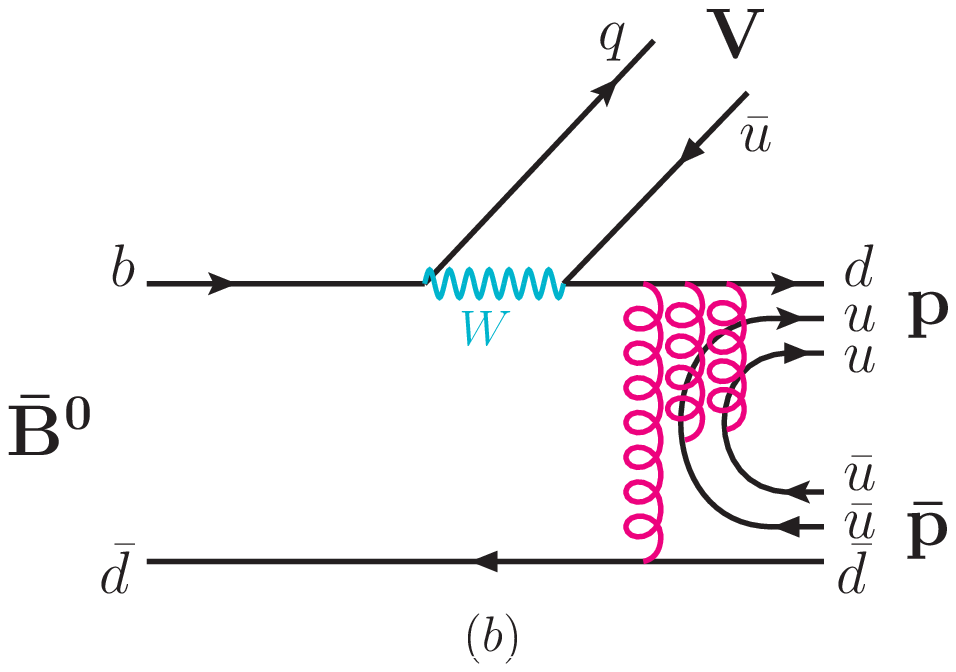}\\
\includegraphics[width=2.2in]{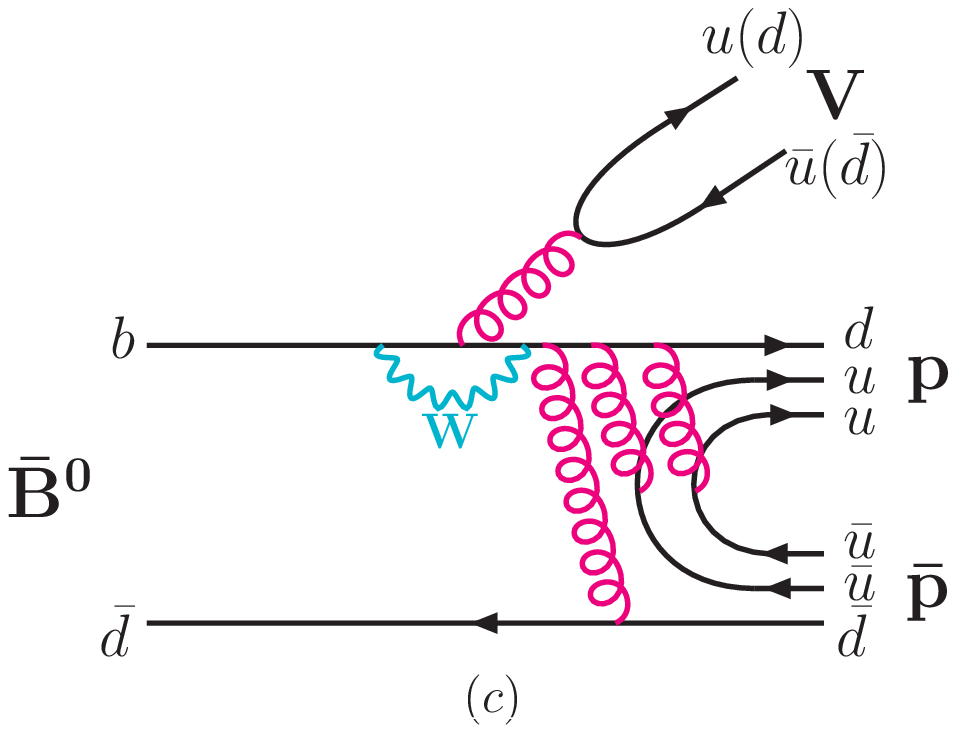}
\includegraphics[width=2.6in]{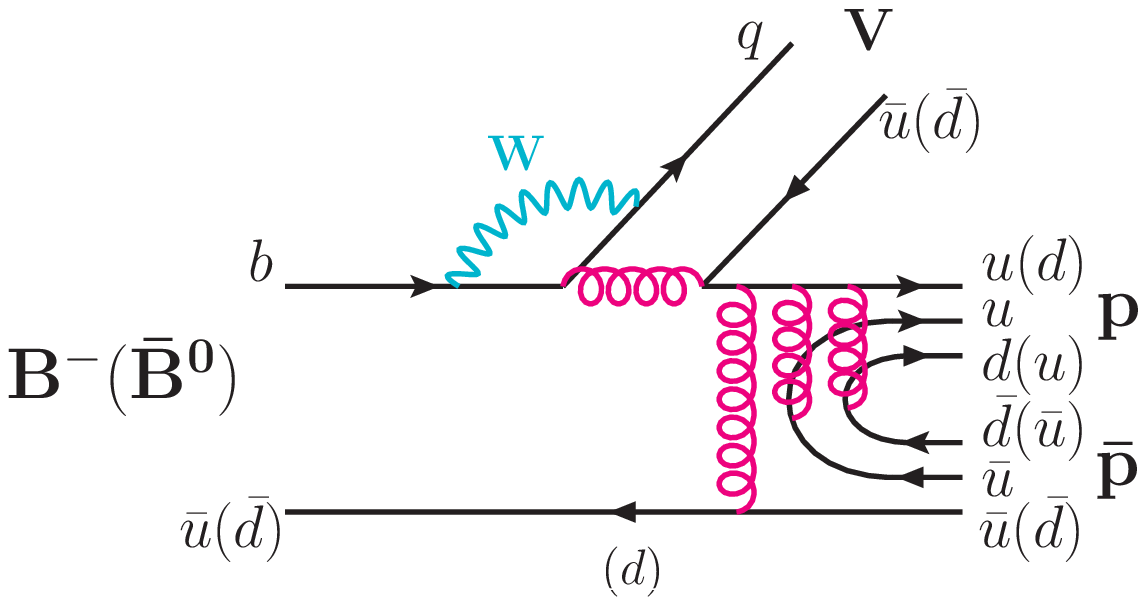}
\caption{Feynman diagrams for $B\to p\bar p V$.}\label{fig2}
\end{figure}
%

To describe the three-body baryonic $B$ decays, 
we take $B\to p\bar p V$ with $V=\rho^{-(0)}$ or $K^{*-}(\bar{K}^{*0})$ as
our examples. According to Fig.~\ref{fig2},
the amplitudes are given by~\cite{Geng:2007cw,Hsiao:2019ann,Geng:2006jt}
\begin{eqnarray}\label{amp1}
{\cal M}(B\to p\bar p V)
&\simeq&\frac{G_F}{\sqrt 2}\alpha_V
\langle V|\bar q\gamma_\mu(1-\gamma_5) q'|0\rangle 
\langle p\bar p|\bar q' \gamma_\mu(1-\gamma_5)b|B\rangle\,,
\end{eqnarray}
with $q=(s,d)$ and $q'=u$ for $B^-\to p\bar p (K^{*-},\rho^-)$, and
$q=(s,d)$ and $q'=d$ for $\bar B^0\to p\bar p (\bar K^{*0},\rho^0)$.
For $\alpha_V$, we define
\begin{eqnarray}\label{amp2}
\alpha_{\rho^-}&=&V_{ub}V_{ud}^* a_1-V_{tb}V_{td}^*a_4\,,\nonumber\\
\alpha_{\rho^0}&=&V_{ub}V_{ud}^* a_2+V_{tb}V_{td}^*(a_4-\frac{3}{2}a_9)\,,\nonumber\\
\alpha_{K^{*-}}&=&V_{ub}V_{us}^* a_1-V_{tb}V_{ts}^*a_4\,,\nonumber\\
\alpha_{\bar K^{*0}}&=&-V_{tb}V_{ts}^*a_4\,.
\end{eqnarray}
In Eq.~(\ref{amp1}), one presents that
$\langle V|\bar q\gamma_\mu(1-\gamma_5) q'|0\rangle
=f_ V m_V \varepsilon_{\mu}^*$, where $f_V$ and $\varepsilon_{\mu}^*$ 
are, respectively, the decay constant and polarization four-vector of the vector meson.
The amplitudes are both associated with 
the matrix elements of the $B\to {\bf B\bar B'}$ transition, and
we parameterize them as~\cite{Geng:2006wz}
\begin{eqnarray}\label{Btoppbar}
\langle {\bf B\bar B'}|V^b_\mu|B\rangle
&=&i\bar u[g_1\gamma_{\mu}+g_2i\sigma_{\mu\nu}p^\nu+
g_3p_{\mu}+g_4(p_{\bf \bar B'}+p_{\bf B})_\mu +g_5(p_{\bf \bar B'}-p_{\bf B})_\mu]\gamma_5v\;,
\nonumber\\
\langle {\bf B\bar B'}|A^b_\mu|B\rangle
&=&i\bar u[f_1\gamma_{\mu}+f_2i\sigma_{\mu\nu}p^\nu+f_3p_{\mu}
+f_4(p_{\bf \bar B'}+p_{\bf B})_\mu +f_5(p_{\bf \bar B'}-p_{\bf B})_\mu]v\;,
\end{eqnarray}
where $V^b_\mu(A^b_\mu)=\bar q' \gamma^\mu(\gamma_5) b$,
$p=p_B-(p_{\bf \bar B'}+p_{\bf B})$, and $(f_i,g_i)$ ($i=1,2, ...,5$) 
are the $B\to {\bf B\bar B'}$ transition form factors. 
Inspired by pQCD counting rules~\cite{Brodsky:1973kr,Brodsky:1974vy,Brodsky:2003gs,
Chua:2002wn,Geng:2006wz}, the momentum dependences for $f_i$ and $g_i$
are given by
\begin{eqnarray}\label{figi}
f_i=\frac{D_{f_i}}{t^n}\,,\;\;g_i=\frac{D_{g_i}}{t^n}\,,
\end{eqnarray}
with $D_{f_i(g_i)}$ to be extracted by the data. According to the gluon lines in Fig.~\ref{fig2},
$n$ in $1/t^n$ should be 2+1, 
which accounts for two gluon propagators attaching to the valence quarks in ${\bf B\bar B'}$
and an additional one for kicking (speeding up) the spectator quark in $B$~\cite{Chua:2002wn}.
For the gluon kicking, it is similar to the meson transition form factor 
derived as $F_M\propto 1/q^2$ in pQCD counting rules~\cite{Lepage:1980fj,Chua:2002pi},
where $1/q^2$ is for a hard gluon 
to transfer the momentum to the spectator quark in the meson.

Like the case of $F_1$ and $g_A$,
we relate $(f_i,g_i)$ to the $B\to{\bf B\bar B'}$ chiral form factors,
which is in terms of~\cite{Chua:2002wn,Geng:2006wz}
\begin{eqnarray}\label{J1J2}
&&
\langle {\bf B}_{R+L}{\bf\bar B'}_{R+L}|(V^b_\mu+A^b_\mu)/2|B\rangle=\nonumber\\
&&
i m_b\bar u\gamma_\mu\bigg[\frac{1+\gamma_5}{2}G_R +\frac{1-\gamma_5}{2}G_L\bigg]u+
i\bar u\gamma_\mu\not{\!p}_b\bigg[\frac{1+\gamma_5}{2}G^j_R +\frac{1-\gamma_5}{2}G^j_L\bigg]u\,,
\end{eqnarray}
where $|B_q\rangle\sim |\bar b\gamma_5 q|0\rangle$ has been used.
In addition, we obtain
$G_{R(L)}=e_{||}^{R(L)} G_{||}+e_{\overline{||}}^{R(L)} G_{\overline{||}}$ and
$G^j_{R(L)}=e_{||}^{R(L)} G^j_{||}+e_{\overline{||}}^{R(L)} G^j_{\overline{||}}$,
which are similar to $F_{R,L}$ in Eq.~(\ref{Gff2}).  
Under the $SU(3)$ flavor and $SU(2)$ spin symmetries,
together with $G_{||(\overline{||})}^{(j)}\equiv D_{||(\overline{||})}^{(j)}/t^n$ $(j=2,3, ...,5)$,
it is derived that~\cite{Chua:2002wn,Geng:2006wz}
\begin{eqnarray}\label{D||}
&&
D_{g_1}=\frac{5}{3}D_{||}-\frac{1}{3}D_{\overline{||}}\,,\;
D_{f_1}=\frac{5}{3}D_{||}+\frac{1}{3}D_{\overline{||}}\,,\;
D_{g_j}=\frac{5}{3}D_{||}^j=-D_{f_j}\,,\nonumber\\
&&
D_{g_1}=\frac{1}{3}D_{||}-\frac{2}{3}D_{\overline{||}}\,,\;
D_{f_1}=\frac{1}{3}D_{||}+\frac{2}{3}D_{\overline{||}}\,,\;
D_{g_j}=\frac{-1}{3}D_{||}^j=-D_{f_j}\,,
\end{eqnarray}
for $\langle p\bar p|(\bar u b)|B^-\rangle$ 
and $\langle p\bar p|(\bar d b)|\bar B^0\rangle$, respectively. 
We also review the direct $CP$ asymmetry, 
defined by
\begin{eqnarray}\label{Acp}
{\cal A}_{CP}(B\to {\bf B\bar B'}M)\equiv
\frac
{\Gamma(B\to {\bf B\bar B'}M)-\Gamma(\bar B\to {\bf \bar B B'}\bar M)}
{\Gamma(B\to {\bf B\bar B'}M)+\Gamma(\bar B\to {\bf \bar B B'}\bar M)}\,,
\end{eqnarray}
where $\Gamma$ denotes the decay width, and
$\bar B\to {\bf \bar B B'}\bar M$ the anti-particle decay.

\section{Numerical Analysis}
For the numerical analysis,  
we adopt the CKM matrix elements as~\cite{pdg}
\begin{eqnarray}
&&
V_{ub}=A\lambda^3(\rho-i\eta)\,,
V_{ud}=1-\lambda^2/2\,,
V_{us}=\lambda,\,\nonumber\\
&&
V_{tb}=1\,,
V_{td}=A\lambda^3\,,
V_{ts}=-A\lambda^2\,,
\end{eqnarray}
with 
$\lambda=0.22453\pm 0.00044$,
$A=0.836\pm 0.015$,
$\bar \rho=0.122^{+0.018}_{-0.017}$,
$\bar \eta=0.355^{+0.012}_{-0.011}$
in the Wolfenstein parameterization,
where $(\bar \rho,\bar \eta)=(1-\lambda^2/2)\times (\rho,\eta)$. 
The decay constants are given by 
$(f_B,f_\rho,f_{K^*})=(0.19,0.21,0.22)$~GeV~\cite{pdg,decayconst}.
With the global fit to the data, 
we obtain~\cite{Hsiao:2014zza,Geng:2016fdw}
\begin{eqnarray}
&&
(C_{||},\,C_{\overline{||}},C_{h_A})=(-102.4\pm 7.3,\,210.9\pm 85.2,8.4\pm 1.4)\,\text{GeV}^{4}\,,\nonumber\\
&&
(D_{||},D_{\overline{||}})=(45.7\pm 33.8,-298.2\pm 34.0)\;{\rm GeV}^{5}\,,\nonumber\\
&&
(D_{||}^2,D_{||}^3,D_{||}^4,D_{||}^5)
=(33.1\pm 30.7,-203.6\pm 133.4,6.5\pm 18.1,-147.1\pm 29.3)\;{\rm GeV}^{4}\,.
\end{eqnarray}
We take $a_i\equiv c^{eff}_i+c^{eff}_{i\pm 1}/N_c$ for $i=$odd (even) with $N_c$
the color number, where the effective Wilson coefficients $c_i$ come from Ref.~\cite{ali}.
For $a_2$ in $\bar B^0\to p\bar p \rho^0$ and $\bar B^0\to p\bar p$,
we use $N_c=2$ to take into account the non-factorizable QCD corrections.
We can hence present our theoretical calculations
for $B\to p\bar p (V)$ in Table~\ref{data2}, along with 
the other $B\to{\bf B\bar B'}M$ results. Besides,
we present our studies of the angular and $CP$ asymmetries.
%
\begin{table}[t!]
\caption{Theoretical results of the two and three-body baryonic $B$ decays,
in comparison with the experimental data.}\label{data2}
{
\tiny
\begin{tabular}{|l|rr|}
\hline
Decay mode              &Exp't data~\cite{pdg,LHCb:2014nix,Belle:2007lbz}&Theory$\;\;\;\;\;\;\;\;\;$\\
\hline\hline
${\cal B}(\bar B^0\to p\bar p)$ 
&$(1.25\pm 0.32)\times 10^{-8}$
&$(1.4\pm 0.5)\times 10^{-8}$~\cite{Hsiao:2014zza}
\\

${\cal B}(\bar B^0\to \Lambda\bar \Lambda)$ 
&$<3.2\times 10^{-7}$
&$(0.3\pm 0.2)\times 10^{-8}$~\cite{Hsiao:2014zza}
\\

${\cal B}(B^-\to \Lambda\bar p)$ 
&$(2.4^{+1.0}_{-0.9})\times 10^{-7}$
&$(3.5^{+0.7}_{-0.5})\times 10^{-8}$~\cite{Hsiao:2014zza}
\\

${\cal B}(\bar B_s^0\to p\bar p)$
& $<1.5\times 10^{-8}$
&$(3.0^{+1.5}_{-1.2})\times 10^{-8}$~\cite{Hsiao:2014zza}
\\
\hline

${\cal B}(\bar B^0\to p\bar p\pi^0)$
&$(5.0\pm 1.9)\times 10^{-7}$
&$(5.0\pm 2.1)\times 10^{-7}$~\cite{Hsiao:2019ann}
\\



${\cal B}(\bar B^0\to \Lambda\bar \Lambda \bar K^{0})$
&$(4.8^{+1.0}_{-0.9})\times 10^{-6}$
&$(2.5\pm 0.3)\times 10^{-6}$~\cite{Geng:2005fh}
\\

${\cal B}(B^-\to p\bar p \pi^{-})$
&$(1.62\pm 0.20)\times 10^{-6}$
&$(1.60\pm 0.18)\times 10^{-6}$~\cite{Hsiao:2015mca}
\\



${\cal B}(B^-\to \Lambda\bar \Lambda \pi^-)$
&$<9.4\times 10^{-7}$
&$(1.7\pm 0.7)\times 10^{-7}$~\cite{Geng:2005fh}
\\

${\cal B}(B^-\to \Lambda\bar \Lambda K^{-})$
&$(3.4\pm 0.6)\times 10^{-6}$
&$(2.8\pm 0.2)\times 10^{-6}$~\cite{Geng:2005fh}
\\

${\cal B}(\bar B^0_s\to \bar p \Lambda K^+$+c.c.)
&$(5.5\pm 1.0)\times 10^{-6}$
&$(5.1\pm 1.1)\times 10^{-6}$~\cite{Geng:2016fdw}
\\

\hline
${\cal B}(\bar B^0\to p\bar p \bar K^{*0})$
&$(1.24^{+0.28}_{-0.25})\times 10^{-6}$
&$(0.9\pm 0.3)\times 10^{-6}$~\cite{Geng:2007cw}
\\

${\cal B}(\bar B^0\to \Lambda\bar \Lambda \bar K^{*0})$
&$(2.5^{+0.9}_{-0.8})\times 10^{-6}$
&$(1.76\pm 0.18)\times 10^{-6}$~\cite{Geng:2011pw}
\\

${\cal B}(B^-\to p\bar p K^{*-})$
&$(3.6^{+0.8}_{-0.7})\times 10^{-6}$
&$(6.0\pm 1.3)\times 10^{-6}$~\cite{Geng:2007cw}
\\

${\cal B}(B^-\to \Lambda\bar p \rho^0,\rho^0\to\pi^+\pi^-)$
&$(4.8\pm 0.9)\times 10^{-6}$
&$(3.28\pm 0.31)\times 10^{-6}$~\cite{Geng:2011pw}
\\

${\cal B}(B^-\to \Lambda\bar p \phi)$
&$(8.0\pm 2.2)\times 10^{-7}$
&$(1.51\pm 0.28)\times 10^{-6}$~\cite{Geng:2011pw}
\\

${\cal B}(B^-\to \Lambda\bar \Lambda K^{*-})$
&$(2.2^{+1.2}_{-0.9})\times 10^{-6}$
&$(1.91\pm 0.20)\times 10^{-6}$~\cite{Geng:2011pw}
\\
\hline
${\cal A}_{FB}(\bar B^0\to \Lambda\bar p \pi^+)$
&$-0.41\pm0.11\pm0.03$
&$(-14.6^{+0.9}_{-1.5}\pm 6.9)\times 10^{-2}$~\cite{Huang:2022oli}
\\
${\cal A}_{FB}(B^-\to \Lambda\bar p \pi^0)$
&$-0.16\pm 0.18\pm 0.03$
&$(-14.6^{+0.9}_{-1.5}\pm 6.9)\times 10^{-2}$~\cite{Huang:2022oli}
\\
${\cal A}_{FB}(B^-\to p\bar p \pi^{-})$
&$-0.409\pm 0.033\pm 0.006$
&$(-49.1^{+0.6}_{-0.3}\pm 1.0\pm 6.3)\times 10^{-2}$~\cite{AFB}
\\
${\cal A}_{FB}(B^-\to p\bar p K^{-})$
&$0.495\pm 0.012\pm 0.007$
&$(46.9^{+0.043}_{-0.041}\pm 0.2\pm 4.7)\times 10^{-2}$~\cite{AFB}
\\
\hline
${\cal A}_{CP}(B^-\to p\bar p \pi^{-})$
&$0.00\pm 0.04$
&$-0.06$~\cite{Geng:2006jt}
\\
${\cal A}_{CP}(B^-\to p\bar p K^{-})$
&$0.00\pm 0.04$
&$0.06\pm 0.01$~\cite{Geng:2006jt}
\\
${\cal A}_{CP}(B^-\to p\bar p K^{*-})$
&$0.21\pm 0.16$
&$0.22\pm 0.04$~\cite{Geng:2006jt}
\\
${\cal A}_{CP}(\bar B^0\to p\bar p\pi^0)$
&
&$(-16.8\pm 5.4)\times 10^{-2}$~\cite{Hsiao:2019ann}
\\
${\cal A}_{CP}(\bar B^0\to p\bar p\rho^0)$
&
&$(-12.6\pm 3.0)\times 10^{-2}$~\cite{Hsiao:2019ann}
\\
\hline
\end{tabular}}
\end{table}

\section{Discussions and Conclusions}
The theoretical results in Table~\ref{data2} can agree well with the data,
which is based on the factorization, pQCD counting rules, and baryonic form factors.
In particular, several $CP$ asymmetries are predicted as large as 10-20\%,
promising to be measured by LHCb and Belle~II~\cite{Belle-II:2018jsg,LHCb:2018roe}.
It is reasonable to extend the theoretical approach to
$B\to{\bf B}_c{\bf\bar B'}$ and $B\to{\bf B\bar B'}M_c$~\cite{Hsiao:2019wyd,Hsiao:2016amt,Chen:2008sw}, 
where ${\bf B}_c(M_c)$ denotes a baryon (meson) containing a charm quark. As a consequence,
$({\cal B},{\cal A}_{FB})$ can also be expalined  (see Table~\ref{data3}).
Nonetheless,
${\cal B}(B^-\to p\bar p\rho^-,\rho^-\to\pi^-\pi^0)=(28.8\pm 2.1)\times 10^{-6}$ and
${\cal B}(B^-\to p\bar p\mu^-\bar \nu_\mu)=(1.04\pm 0.24\pm 0.12)\times 10^{-4}$
we have predicted are not verified by the observations~\cite{Geng:2007cw,
Geng:2011tr,pdg,LHCb:2019cgl},
\begin{eqnarray}\label{BtoppX}
&&{\cal B}(B^-\to p\bar p\pi^-\pi^0)
=(4.6\pm 1.3)\times 10^{-6}\,,\nonumber\\
&&{\cal B}(B^-\to p\bar p\mu^-\bar \nu_\mu)
=(5.27^{+0.23}_{-0.24}\pm 0.21\pm 0.15)\times 10^{-6}\,,
\end{eqnarray}
where the amplitude of $B^-\to p\bar p\mu^-\bar \nu_\mu$
is given by
\begin{eqnarray}\label{semi}
{\cal M}(B^-\to p\bar p \ell \bar \nu_\ell)
&=&\frac{G_F}{\sqrt 2}V_{ub}
\langle p\bar p|\bar u\gamma_\mu (1-\gamma_5)b|B^-\rangle 
\bar \ell\gamma^\mu (1-\gamma_5) \nu_\ell\,.
\end{eqnarray}
Since $B\to p\bar p\rho$ and $B^-\to p\bar p\mu^-\bar \nu_\mu$
are seen to be associated with the $B\to p\bar p$ transition form factors,
which are inferred to cause the overestimations.
Besides, ${\cal B}(B^-\to p\bar p\mu^-\bar \nu_\mu)$
inconsistent with the data can be partly due to the inconsistent determination of $|V_{ub}|$
between the inclusive and exclusive $B$ decays.

As the final remark, 
since the predictions of ${\cal B}(B^-\to p\bar p\rho^-)$ and 
${\cal B}(B^-\to p\bar p\mu^-\bar \nu_\mu)$ are shown to deviate from
the observations by the factors of 6 and 20, respectively, 
the theoretical approach is facing some difficulties.
Therefore, the re-examination should be performed elsewhere.
%
\begin{table}[t!]
\caption{Theoretical results of the two and three-body $B$ decays
with the baryon (meson) containing a charm quark, 
which are compared with the experimental data.}\label{data3}
{
\tiny
\begin{tabular}{|l|rr|}
\hline
Decay mode              &Exp't data~\cite{pdg,Chang:2015fja}$\;\;\;$&Theory$\;\;\;\;\;\;\;\;\;$\\
\hline\hline
${\cal B}(\bar B^0\to \Lambda_c^+ \bar p)$
&$(1.54\pm 0.18)\times 10^{-5}$
&$(1.0^{+0.4}_{-0.3})\times 10^{-5}$~\cite{Hsiao:2019wyd}\\
${\cal B}(\bar B^0\to \Sigma_c^+ \bar p)$
&$<2.4\times 10^{-5}$
&$(2.9^{+0.8}_{-0.9})\times 10^{-6}$~\cite{Hsiao:2019wyd}\\
\hline
${\cal B}(B^-\to \Lambda\bar p D^0)$
&$(1.43\pm 0.32)\times 10^{-5}$
&$(1.14\pm 0.26)\times 10^{-5}$~\cite{Chen:2008sw}\\
${\cal B}(B^-\to \Lambda\bar p D^{*0})$
&$<5\times 10^{-5}$
&$(3.23\pm 0.32)\times 10^{-5}$~\cite{Chen:2008sw}\\
${\cal B}(\bar B^0\to n\bar p D^{*+})$
&$(1.4\pm 0.4)\times 10^{-3}$
&$(1.45\pm 0.14)\times 10^{-3}$~\cite{Chen:2008sw}\\

${\cal B}(\bar B^0\to p\bar p D^{0})$
&$(1.04\pm 0.07)\times 10^{-4}$
&$(1.04\pm 0.12)\times 10^{-4}$~\cite{Hsiao:2016amt}\\
${\cal B}(\bar B^0\to p\bar p D^{*0})$
&$(0.99\pm 0.11)\times 10^{-4}$
&$(0.99\pm 0.09)\times 10^{-4}$~\cite{Hsiao:2016amt}\\
${\cal B}(\bar B^0\to \Lambda\bar p D^+)$
&$(2.5\pm 0.4)\times 10^{-5}$
& $(1.85\pm 0.30)\times 10^{-5}$~\cite{Hsiao:2016amt}\\

${\cal B}(\bar B^0\to \Lambda\bar p D^{*+})$
&$(3.4\pm 0.8)\times 10^{-5}$
&$(2.75\pm 0.24)\times 10^{-5}$~\cite{Hsiao:2016amt}\\
${\cal B}(\bar B^0\to \Sigma^0\bar \Lambda \bar D^0+c.c.)$
&$<3.1\times 10^{-5}$
&$(1.8\pm 0.5)\times 10^{-5}$~\cite{Chen:2008sw}\\
\hline
${\cal A}_{FB}(\bar B^0\to \Lambda\bar p D^+)$
&$-0.08\pm 0.10$
&$-0.030\pm 0.002$~\cite{Hsiao:2016amt}\\

${\cal A}_{FB}(\bar B^0\to \Lambda\bar p D^{*+})$
&$+0.55\pm 0.17$
&$+0.150\pm 0.000$~\cite{Hsiao:2016amt}
\\

\hline
\end{tabular}}
\end{table}

\newpage
In summary, to review the baryonic $B$ decays,
we have summarized the experimental data,
which includes branching fractions, angular and $CP$ asymmetries.
We have taken $\bar B^0\to p\bar p$ and  
$B\to p\bar p V$ with $V=\rho^{-(0)}$ or $K^{*-}(\bar{K}^{*0})$ for theoretical illustration.
We have also reviewed the $CP$ asymmetries of $B\to p\bar p M$, 
which can be used to compare with future measurements by LHCb and Belle~II.
With the theoretical results listed in the tables, we have demonstrated that
the theoretical approach can be used to interpret most observations. 
Finally, we have also remarked that 
the theoretical approach has currently encountered some challenges in
interpreting ${\cal B}(B^-\to p\bar p\rho^-)$ and 
${\cal B}(B^-\to p\bar p\mu^-\bar \nu_\mu)$.

\section*{ACKNOWLEDGMENTS}
The authors would like to thank 
Professors~Xin Liu, Zhen-Jun Xiao, Qin Chang, and Rui-Lin~Zhu
for inviting us to write a review article on the physics of baryonic B meson decays 
in the special issue ``Heavy Flavor Physics and CP Violation'' of Advances in High Energy Physics.
YKH was supported in part by NSFC (Grant Nos.~11675030 and 12175128).
LS was supported in part by NSFC (Grant No.~12061141006)
and Joint Large-Scale Scientific Facility Funds of the NSFC and CAS (Grant No.~U1932108).


\end{document}